# Optimization of a NdFeB permanent magnet configuration for in-vivo drug delivery experiments

A.Omelyanchik[1,4], G.Lamura[2], D.Peddis[3,4], and F.Canepa[2,4]*

[1] Immanuel Kant Baltic Federal University, Kaliningrad, Russian Federation
[2] CNR-SPIN, Corso Perrone 24, I-16152, Genova, Italy
[3] Istituto di Struttura della Materia – CNR, Monterotondo Scalo, RM, Italy
[4] Department of Chemistry and Industrial Chemistry Università di Genova, Genova, Italy

## Abstract

We propose a new concept of magnetic focusing for targeting and accumulation of functionalized superparamagnetic nanoparticles in living organs through composite configurations of different permanent magnets. The proposed setups fulfill two fundamental requirements for *in vivo* experiments: 1) reduced size of the magnets to best focusing on small areas representing the targeted organs of mice and rats and 2) maximization of the magnetic driving force acting on the magnetic nanoparticles dispersed in blood. To this aim, several configurations of permanent magnets organized with different degrees of symmetry have been tested. The product $B \cdot grad(B)$ proportional to the magnetic force has been experimentally measured, over a wide area (20× 20 mm²), at a distance corresponding to the hypothetical distance of the mouse organ from the magnets. A non-symmetric configuration of mixed shape permanent magnets resulted in particularly promising to achieve the best performances for further *in vivo* experiments.

* Corresponding author.

E-mail address: fabio.canepa@unige.it

# 1. Introduction

In the last years, functionalized monodomain Magnetic NanoParticles (MNPs) as carriers for magnetic drug delivery therapy have been extensively studied as a promising less destructive alternative to typical chemiotherapic protocols against several types of cancer [1–3]. Special attention was given to side-specific targeting of stem cells enhanced with gene therapy [4–6] which invoked interests in the fundamental studies of influences of the magnetic field and MNPs on the viability and manipulation of the cell`s behavior [7–9]. This approach is very promising for example for engraftment of cells of the cardiovascular system after surgery.

However, while exhaustive researches on the MNPs synthesis and functionalization for drug delivery have been carried out [10,11], relatively few studies have been performed on the optimal magnetic field and magnetic gradient parameters required to transport and accumulate the nanoparticles in the targeted organ [12]. While the achievement of the high-value gradient magnetic field at low-dimension is the relatively well-developed field [8,13,14], the upscaling of those configurations to the scale of human organs is a complicated task because of the fast attenuation with the increase of the distance.

Typically, the production of a magnetic field can be achieved by the use of 1) a superconducting magnet, 2) an electromagnet or 3) a configuration of permanent magnets. Some results concerning the development of a system for magnetic drug delivery have been performed by the use of a superconducting magnet [15,16], for example, a superconducting magnet of magnetic resonance imaging device [17]. However, even if its use allows to obtain very high values of the magnetic field and also to its gradient in the flow direction, the complex experimental configuration and the quite expensive maintenance costs, prevent the use of a superconducting magnet for *in vivo* experiments. Electromagnetic systems for magnetic drug delivery were mainly developed from a theoretical point of view, by 3D designs and simulation methods, using appropriate software as Comsol Multiphysics [18] or Finite Element Model (FEM) [19].

At present, the most feasible way to obtain a useful magnetic field for *in vitro* and *in vivo* drug delivery experiments is the use of a single or a suitable configuration of permanent magnets. On one side, theoretical simulations have been performed on an ideal system of MNPs dispersed in blood using different configurations of permanent magnets [20–25]. On the other side, different *in vitro* experiments have been carried out using a single magnet or a configuration of multiple magnets separated by non-magnetic materials [26]. In several studies *in vivo* experiments of magnetic drug delivery and targeting to a specific organ have been performed using a suitable configuration of NdFeB permanent magnets [27–31]. For example, the configuration of permanent magnets was used to radially symmetric cell deposition in vessels of mice [6,32]. In those studies, combined action magnetic nanoparticles and fields reinforced gene and cell therapy of vessels after irreversible damage caused for example by mechanical denudation. In most of the related works, attention was devoted to the magnetic field produced by them, but not to the gradient required or the magnetic driving force (proportional to their product $B \cdot grad(B)$) to focalize the functionalized MNPs to the specific organ and no attention was devoted to the relative dimensions of the magnets with respect to the targeted organ as well.

So, there is a lack of information regarding the choice of the permanent magnets geometry and their relative dimensions with respect to the mice used for *in vivo* tests and the direct 2D mapping of the magnetic field of different permanent magnets configurations obtained at different heights as well as calculations concerning the relative gradients and the magnetic force.

Since the *in vivo* tests require a lot of time (from 2 up to 4 hours) in order to ensure the accumulation of functionalized MNPs in the targeted organ, the small dimensions of the rats used for the experiments prevent the use of large (and heavy) permanent magnets. So, the utilization of a light system of small permanent magnets in a non-metallic thin structure can significantly increase the test efficiency.

Therefore, this paper aims to present a preliminary experimental study on three different permanent magnets configurations useful for *in vivo* magnetic drug delivery, compared to the results achieved by the use of a single magnet with the same grade. The target is to maximize the magnetic driving force over a surface area of about 1 cm², (*i.e.* a value comparable to the typical surface of an organ of a mouse as hearth, liver or lungs). The studied configurations of magnets were designed for *in vivo* experiments where magnets will be placed on the skin of mice without surgery to enhance the accumulation of MNPs in a specific organ. The magnetic mapping was performed by the direct measurement of the magnetic induction on the surface of the system and at the distance of Z = 4 mm from it, *i.e.* the typical distance between a mouse organ and the skin. A configuration that creates the strongest product in regions of interest was selected further in vivo experiment.

## 2. Experimental details

All the commercial $Nd_2Fe_{14}B$ permanent magnets were from HKCM Engineering e.K. (DE), with the same N52 quality grade. For the magnetic configurations, two different magnets geometries were used: a cubic geometry with l = 5 mm and a cylindrical one with 3×3 mm size (diameter × length), with easy magnetization direction axially oriented. All magnets were Ni coated. Detailed characterization of the magnetic properties of cylindrical magnets of the same grade was reported in our previous work [33]. At room temperature remanent magnetic induction of single magnet was maximal among a set of commercially available magnets and reached about 14 kiloGauss (kG).

For the magnetic induction ($\vec{B}$) measurements, a Lake Shore gaussmeter model 475 DSP coupled with an axial Lake Shore 400 HSE Hall probe (Lake Shore Cryotronics Inc., USA) was used. Thus the instrument gives the measurement of the axial component of the induction field $\vec{B}$ vector. The sensitivity of the probe was better than 5 milli Gauss (mG), while the magnetic induction range was up to 35 kG. A nominal active area of sensing element reported by the manufacturer was about of 1 mm in diameter. The stem for Hall probe was made from fiberglass epoxy and had a diameter of 5 mm. The precise positioning of the Hall probe was achieved by a homemade positioner setup with a 3-axial coordination with the error of 0.05 mm. The stem was therefore fixed vertically respect to the magnet plane. By considering a coordinate system x,y,z with its origin on the magnets flat surfaces, the value returned by the instrument is the z component of the induction field as a function of the in-plane coordinates at a fixed $\bar{z}$ height $B_{measured} = B_z(x, y, \bar{z})$.

At first step, the magnetic induction produced by a single cylindrical magnet (diameter and height equal to 3 mm) and by different configurations of several magnets were mapped by measuring $B_z(x, y, \bar{z})$ over a 10×10 mm² surface by 1 mm step at fixed height $\bar{z}$ = 4 mm from the magnet surface. At each x-y step of the magnetic induction map, the in-plane gradient components were calculated with the following equations:

$$\begin{cases} \left(grad(B_z(x,y,\bar{z}))\right)_x = \left(\frac{\Delta B_z(x,y,z)}{\Delta x}\right)_{y,z=cost.} \\ \left(grad(B_z(x,y,\bar{z}))\right)_y = \left(\frac{\Delta B_z(x,y,z)}{\Delta y}\right)_{x,z=cost.} \end{cases}.$$

Besides singe magnet, three magnetic configurations were adopted:

a) Three cylindrical magnets were inserted in a Teflon disk of 13 mm diameter, 3 mm thickness, used as support, at the vertices of an equilateral triangle of 4 mm side. All the magnets presented the same polarization direction (Fig.1A).
b) Four cylindrical magnets inserted in a squared Teflon disk (20 mm side, 3 mm thickness) at the vertices of a square of 5 mm side. Yet all the magnets were set with the same polarization direction (Fig.1B).
c) A mixed configuration fixed again in a squared Teflon disk (20 mm side, 5 mm thickness): two cubic magnets at two vertices of an ideal square of 5 mm side on the same diagonal and six cylindrical magnets, three by three, fixed in the remanent two vertices along the other diagonal. In this case, the polarity of magnets was changed starting from mutually opposite polarization of cube and cylinder magnets to the same orientation of all magnets (see Fig.1C).

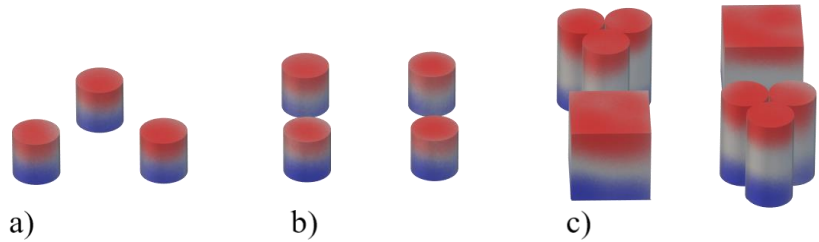

a)  b)  c)

*Figure 1.* Visualization three cylindrical magnet system (a), four cylindrical magnet system (b), and mixed system of cubic and cylinder magnets (c).

In all cases, the weight of the magnet set-up is less than 10 grams in order to prevent any physical impediment to the rats and to allow an easy fixing to the selected position during *in vivo* experiments.

## 3. Results and discussion

### 3.1. Magnetic assumptions

The magnetization of MNPs in the superparamagnetic limit at low field is a linear function of the applied magnetic field as $M = \chi_{NP} B$, where $\chi_{NP}$ is the initial mass magnetic susceptibility [34]. Roughly, in the general case of iron based MNPs (e.g. spin iron oxide, iron) with size below 10 nm (the most studied case for biomedical applications) this is largely verified for fields below 10 kGauss. Thus, in field range considered in this study (hundreds of gausses) a MNP with mass $m_{NP}$ in a dynamic fluid with fixed flow rate, is subjected to a magnetic force $(F_M)_{x,y}$ whose in-plane components $(F_M)_x$ and $(F_M)_y$ are defined here below:

$$(F_M)_{x,y} = m_{NP} \chi_{NP} B_z(x,y,\bar{z}) \cdot [grad(B_z(x,y,\bar{z}))]_{x,y} \quad (1)$$

where, $B_z(x, y, \bar{z})$ is the value of the z component of the magnetic induction at a fixed distance $\bar{z}$ from the magnet and $grad(B_z(x,y,\bar{z})) = \sqrt{[grad(B_z(x,y,\bar{z}))]_x^2 + [grad(B_z(x,y,\bar{z}))]_y^2}$ is the intensity of in-plane gradient of $B_z$ measured at a fixed $\bar{z}$. In the rest of the article we will refer simply to $B \cdot grad(B)$. Typically, the diamagnetic contribution of the fluid can be neglected, as well as that ascribed to the organic coating. So, it is evident that the request of a strong force acting on a MNP needs a large value of the product $\cdot grad(B)$. This means that not only $B$ must be the maximum at the $\bar{z}$ distance from the target, but also its gradient along $x$ and $y$ directions. For this very reason, in the following paragraphs we will present the measurements of the induction field, the in-plane gradient and their products for several magnet configurations.

### 3.2. Single cylindrical magnet

The measurements of the magnetic induction obtained, at $\bar{z} = 4$ mm, for a single cylindrical magnet are reported in Fig.2A. In Fig.2B, we report the calculated gradient for each step (1 mm) of the 2D surface map displayed in Fig.2A. The length of arrows is proportional to the magnitude of gradient while the orientation of arrows shows the direction to increase of the magnetic induction magnitude. We subsequently calculated $B \cdot grad(B)$ map as presented in Fig.2C, where the color grade represents the intensity of the magnetic force acting on MNPs. In particular, the dark red circular crown represents the area where the maximum accumulation of MNPs is expected to be localized.

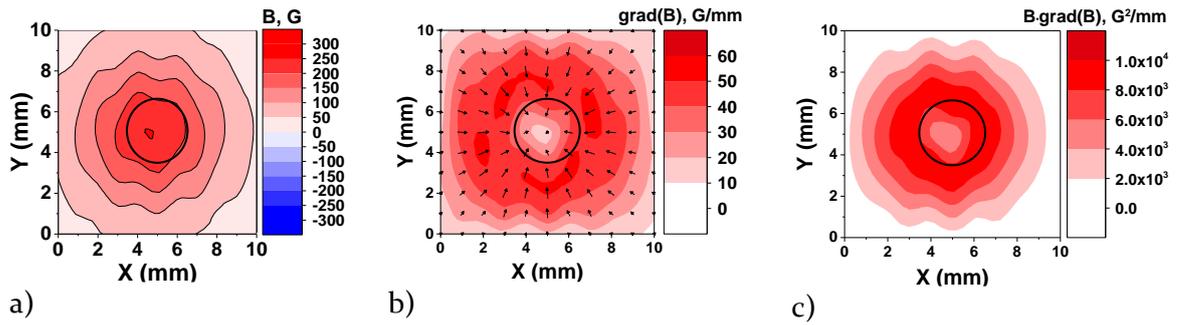

*Figure 2.* Map of magnetic induction (a), the gradient of magnetic induction (b) and value of the product (c) for a single cylindrical magnet at the surface under the magnet ($\bar{z} = 4$ mm). The black line represents the magnet's real dimension and shape.

### 3.3. Configurations of three and four cylindrical magnets

#### 3.3.1. Three cylindrical magnets

An improvement of the attracting force acting on the MNP (see eq. 1) can be achieved increasing the product $B \cdot grad(B)$ playing with the spatial magnetic induction distribution generated by several magnets such as the three-magnet set-up. This configuration has also the advantage to increase the surface area on which such product is effective. The 2D-map of· for the present configuration at fixed height $\bar{z} = 4$ mm is presented in Fig.3A. To estimate the spatial distribution of the product, we calculated $B \cdot grad(B)$ averaged over a 10×10 mm² Region Of Interest (ROI). The value of the product averaged over ROI and the maximum value of the product achieved in a single point are reported in Table 1. In order to maximize the induction gradient, we considered two magnets UP and one DOWN set-up as well. The product $B \cdot grad(B)$ was less satisfactory for this configuration and therefore they are not presented here.

#### 3.3.2. Four cylindrical magnets

In Fig.3B, the magnetic pattern (at $\bar{z} = 4$ mm) of the system of four magnets distributed in corners of the Teflon square is presented. Also, in this case, an alternative configuration, formed by coupled UP and DOWN magnets along the diagonals of the square was taken into account. As in the previous case of three magnets system, a reduced value of the product $B \cdot grad(B)$ were obtained (not shown). Notable that the symmetrical configuration of magnets demonstrates a slightly asymmetric pattern of the magnetic gradient probably because of the different properties of each individual magnets even from the same bunch.

A comparison of the triangular and squared set-ups demonstrates that $B \cdot grad(B)$ is higher over one single spot in the former, while in the latter it results distributed over longer distances, thus allowing to achieve a higher area where the magnetic force can attract MNPs. Also, for the four magnets configuration, the best averaged product over ROI and the maximum value in a single point are reported in the same Table 1.

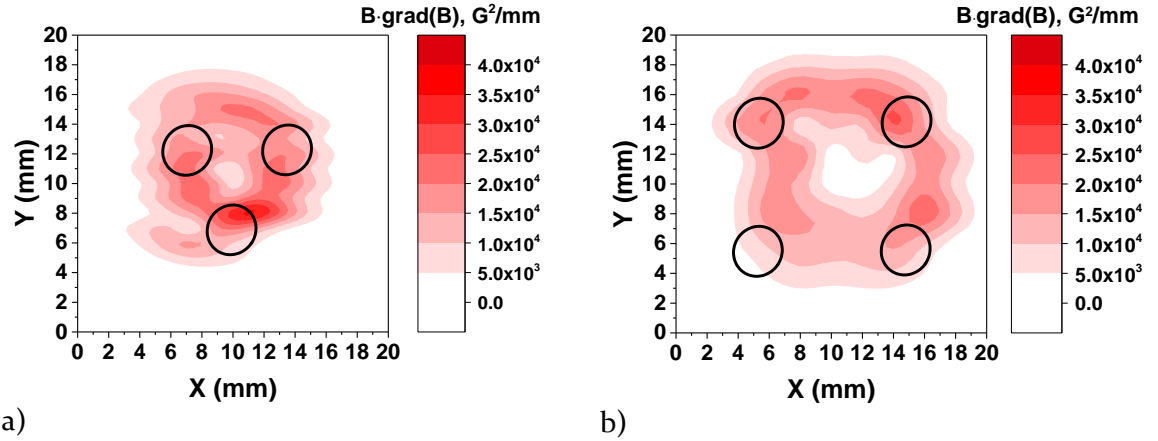

a) b)

***Figure 3.*** *Map of the product for systems of three (a) and four (b) cylindrical magnets all up at the surface under the magnet ($\bar{z}$ = 4 mm). The black lines represent the magnet's real dimensions and shapes.*

### 3.4. Configurations of mixed systems of cylindrical and cubic magnets

In the above configurations of small cylindrical magnets (3×3 mm), the contribution of $grad(B)$ to the magnetic force of eq. (1) is small and therefore $F_M$ is mainly due to the magnetic induction $B$. Therefore, for further improvement of $F_M$, a combination of cylindrical and cubic magnets was designed and realized: the results concerning the level $\bar{z}$ = 4 mm are reported in Fig.4. We measured all the possible configurations resulting by changing the magnet polarity at that fixed distance. In Fig.4A the $B \cdot grad(B)$ 2D-map is plotted when the polarity of cubic magnets was opposite to cylindrical ones. In such a case opposite oriented magnets suppress the induction of each other: as a result, this configuration exhibits a low value of B and consequently $B \cdot grad(B)$. Turning one of the sextets of cylindrical magnets, a higher value of $F_M$ was achieved (Fig.4B). A similar good result was obtained also with all magnets polarized in the same direction (Fig.4C). But in this case, in the center of the system, $B$ has a plateau and thus $grad(B)$ assumes very low intensities thus decreasing consequently $F_M$ as well. We note that the value of ROI parameter was smaller than what was obtained when one of the sextets magnet groups was turned opposite (Fig.4B and Table 1).

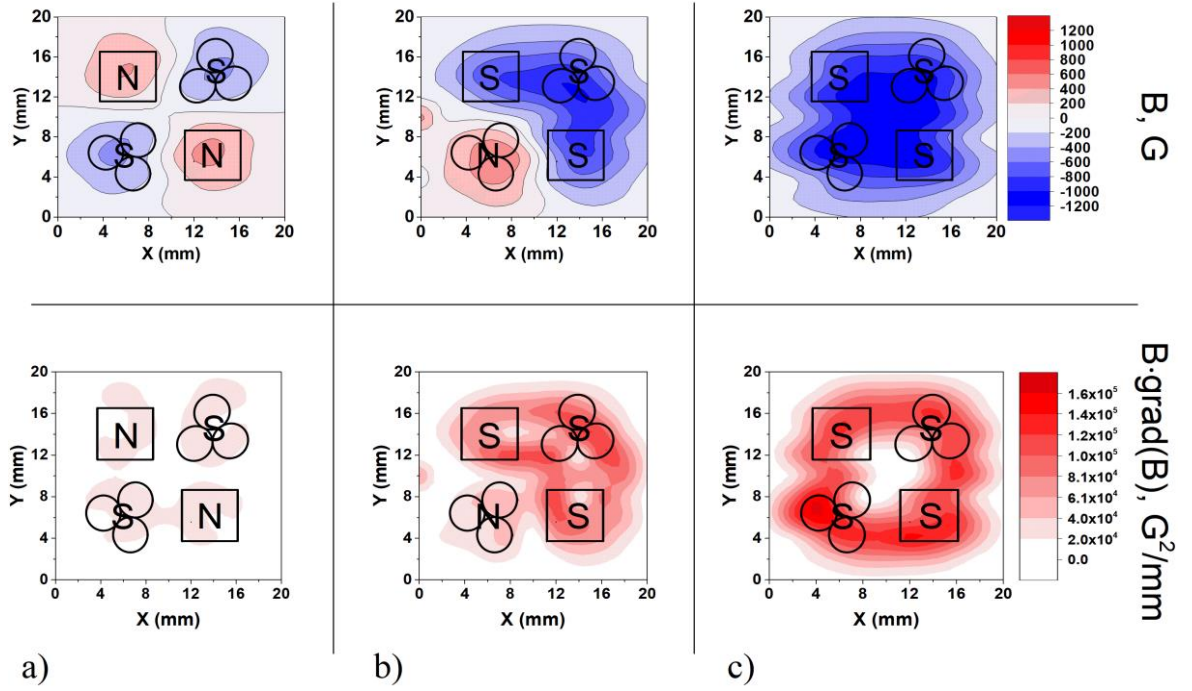

a)   b)   c)

**Figure 4.** *2D-Map of B (upper panel) and $B \cdot grad(B)$ (bottom panel) at $\bar{z}$ = 4 mm for systems of mixed cubic and cylindrical magnets with the opposite polarity of cubic magnets to cylindrical ones (a), one of the sextets of cylindrical magnets with opposite polarity (b) and all magnets polarized in the same direction (c). The black lines represent the magnet's real dimensions and shapes.*

**Table 1.** *Maximum value of $B \cdot grad(B)$ and its average value over ROI at $\bar{z}$ = 4 mm.*

| Configuration | $B \cdot \mathbf{grad}(B)$, G²/mm | |
|---|---|---|
| | Maximum value in a single point | Average over ROI of 10×10 mm² square |
| 3 cylinder magnets (Fig.3A) | 3.7 10⁴ | 1.7 10⁴ |
| 4 cylinder magnets (Fig.3B) | 2.7 10⁴ | 2.0 10⁴ |
| Mixed 2UP/2DN (Fig.4A) | 3.7 10⁴ | 2.9 10⁴ |
| Mixed 3UP/1DN (Fig.4B) | 1.2 10⁵ | 9.7 10⁴ |
| Mixed 4UP (Fig.4C) | 1.6 10⁵ | 8.1 10⁴ |

From the data reported in Table 1, adopting the configuration of the magnets reported in Fig.4B, the average value of the product $B \cdot grad(B) = 9.7 \times 10^4$ G²/mm can be obtained over an area of 10×10 mm². Mouse heart dimensions are about 10×4.2 mm² [35], while mouse liver presents a larger volume, around 22 cm³ [36]. These values suggest that the above magnets configurations can be successfully used for *in vivo* experiments of magnetic drug delivery and targeting.

## Conclusions

Despite that the physics behind the propagation of magnetic fields in the space is well known, the way of the shaping of magnets or their systems to satisfy the requirements of certain bio-application is relatively missed in literature. Here the new insights on the magnetic profile produced by different configurations of small permanent magnets are presented. The choice of dimensions and the adopted geometries are strictly related to the dimensions of the mice organs typically used in *in vivo* experiments. Our results give evidence that the magnetic force is maximized by a subtle balance between the polarity configuration and the geometric shape of the used magnets. In particular, we found that, over the typical mice organ size, the average product of $B \cdot grad(B)$ attains $10^5$ G²/mm with a mixed magnets configuration with one of the cylindrical magnet groups with opposite polarity respect to the others. *In vivo* experiments are in progress in order to validate our experimental achievements.

**Declaration of Competing Interest**
The authors declare that they have no known competing financial interests or personal relationships that could have appeared to influence the work reported in this paper.

**CRediT authorship contribution statement**
Alexander Omelyanchik: Visualization, Investigation, Writing- Original draft preparation; Gianrico Lamura: Investigation, Methodology, Conceptualization, Writing- Original draft preparation, Writing- Reviewing and Editing, Data curation; Davide Peddis: Writing- Reviewing and Editing, Conceptualization, Methodology, Data curation, Supervision; Fabio Canepa: Writing- Reviewing and Editing, Conceptualization, Methodology, Data curation, Supervision.